\documentclass[trackchanges,twocolumn]{aastex701}

\newcommand{\elow}{$E_\mathrm{low}$}
\newcommand{\loggf}{$\log(gf)$}

\usepackage{subcaption}

\begin{document}

\title{Metallicity does not reorder line-to-line sensitivity to granulation in solar-like stars}

\correspondingauthor{D.~Vukadinovi\'{c}}
\email{dusan.vukadinovic@uni-graz.at}

\author[orcid=0000-0003-1971-5551,sname='Vukadinovi\'{c}']{D.~Vukadinovi\'{c}}
\affiliation{Institute of Physics, University of Graz, Universit\"{a}tplatz 5, 8010, Graz, Austria}
\email[]{dusan.vukadinovic@uni-graz.at}  

\author[orcid=0000-0002-3243-1230]{K.~Sowmya}
\affiliation{Institute of Physics, University of Graz, Universit\"{a}tplatz 5, 8010, Graz, Austria}
\email[]{sowmya.krishnamurthy@uni-graz.at}

\author[orcid=0009-0009-3020-3435]{V.~Vasilyev}
\affiliation{Max-Planck-Institut f\"ur Sonnensystemforschung, Justus-von-Liebig-Weg 3, 37077 G\"ottingen, Germany}
\email[]{Vasilyev@mps.mpg.de}

\author[orcid=0000-0002-0929-1612]{V.~Witzke}
\affiliation{Institute of Physics, University of Graz, Universit\"{a}tplatz 5, 8010, Graz, Austria}
\email[]{veronika.witzke@uni-graz.at}

\author[0000-0002-8863-7828]{A.~Collier~Cameron}
\affiliation{Centre for Exoplanet Science, SUPA School of Physics and Astronomy, University of St Andrews, North Haugh, St Andrews KY16 9SS, UK}
\email[]{andrew@standrews.uk}

\begin{abstract}

Stellar granulation imprints line-dependent wavelength shifts and profile variations that limit the precision of radial velocity (RV) measurements used for exoplanet detection and characterization. Understanding how the sensitivity of individual spectral lines to granulation depends on stellar parameters is therefore essential for mitigating granulation-induced RV signals. Here, we investigate the effect of metallicity using MURaM 3D magnetohydrodynamic simulations of stellar granulation including a small-scale dynamo for metallicities $\mathrm{M/H}=\{-1.0,-0.5,0.0,0.3,0.5\}$. We synthesize spectra of neutral and ionized lines spanning a broad range of lower excitation potentials and transition probabilities using the MPS-ATLAS code, assuming local thermodynamic equilibrium. We quantify the sensitivity to granulation using the continuum-intensity-weighted standard deviations of the line centre-of-gravity wavelength and equivalent width, denoted by $\sigma_\lambda$ and $\sigma_W$, respectively. Both metrics increase systematically with metallicity at disc centre owing to the combined effects of more vigorous convection and metallicity-dependent changes in the line opacity. We find that the correlation between the solar ($\mathrm{M/H}=0.0$) and non-solar sensitivity metrics remains high from the disk center to the stellar limb. This indicates that the relative ordering of line sensitivities is robust against changes in viewing angle and metallicity. Consequently, line masks calibrated on the solar spectrum can be reliably applied to radial-velocity measurements of stars with non-solar metallicities, provided they have solar effective temperature and surface gravity.

\end{abstract}

\keywords{Stellar photosphere -- Stellar granulation -- Radiative transfer -- Metallicity -- Radial velocity -- Exoplanets}


\section{Introduction}

The detection and characterization of Earth-like exoplanets orbiting Sun-like stars requires radial velocity (RV) measurements with precisions on the order of 10\,cm\,s$^{-1}$ \citep{Fisher16}. New generation spectrographs such as ESPRESSO \citep{Pepeetal2021}, NEID \citep{Schwabetal2016}, and EXPRES \citep{Jurgensonetal2016}, are approaching this level of instrumental stability. At these precisions, stellar signals become the dominant source of noise for detecting Earth-like exoplanets. Flows associated with surface inhomogeneities such as granulation and magnetic activity, together with stellar oscillations, produce RV variations that can mask or mimic planetary signals \citep[see e.g.][and references therein]{Meunieretal2010,Fisher16,Crassetal2021}. Consequently, the detection of Earth analogs is no longer limited primarily by instrumental capabilities, but by variability intrinsic to the host star. Improving the extraction of the planet-induced RV signal therefore requires a detailed understanding of the spectral signatures of stellar variability and the development of effective strategies to distinguish stellar signals from true planet-induced Doppler shifts.

The most widely used approaches for measuring high-precision RVs are the cross-correlation function (CCF) method and template matching. In the CCF approach, the observed spectra are cross-correlated with carefully constructed numerical masks \citep[i.e., line mask;][]{Baranne96} with the heavier weighting of deep and sharp lines, which are the most sensitive ones to the velocity changes \citep{Pepe02}. Template-matching methods instead determine the wavelength shift that best aligns an observed spectrum with a high signal-to-noise reference template constructed from the observations themselves \citep[e.g.,][]{Anglada-Escude12}. Despite their methodological differences, both approaches derive a single RV measurement by combining information from a large number of spectral lines distributed across the stellar spectrum. This implicitly assumes that the selected spectral lines respond similarly to stellar variability and can therefore be combined to provide a robust estimate of the planet-induced RV.

Spectral line response to intrinsic stellar variability is, however, strongly differential. The planetary reflex motion produces a common Doppler shift across the stellar spectrum, whereas stellar variability introduces line-dependent changes in line depths, shapes, and asymmetries in addition to apparent Doppler shifts \citep{Dravinsetal1981,Dravins1982,Dravins1987a,Dravins1987b,Gray1981,Gray1982,Gray1983,Gray1989}. Characterizing these line-dependent signatures is therefore key to overcoming the stellar variability barrier. To this end, a wide range of mitigation strategies has been developed that exploit the different ways stellar variability manifests itself in spectral lines. For example, the CCF-based SCALPELS algorithm \citep{CollierCameron21} uses a mitigation strategy based on separating shift-like RV signals from those driven by changes in shape and relative strengths of spectral lines. A complementary approach is provided by line-by-line methods, which identify and preferentially weight spectral lines according to their sensitivity to stellar variability \citep{Dumusque18,Cretignier20,AlMoulla22}. At present, these sensitivities are typically inferred empirically from observations or estimated using simplified theoretical considerations. Physics-based predictions of line sensitivities across stellar parameter space and over the full wavelength range covered by modern RV spectrographs would substantially enhance the predictive power of these approaches.

Physics-based predictions of spectral-line sensitivities have recently become feasible owing to major advances in three-dimensional (magneto)hydrodynamic ((M)HD) simulations of stellar photospheres, which successfully reproduce a wide range of observed properties of stellar convection and spectral line formation \citep[see e.g.,][]{Witzke24}. In the context of granulation, \citet{Ceglaetal2013,Ceglaetal2018,Ceglaetal2019,Frameetal2025} used 3D MHD simulations computed with the MURaM code \citep{Voegleretal2004,Voegleretal2005,Rempel14} to investigate granulation-induced temporal signatures in selected spectral lines. Using HD simulations generated with the CO5BOLD code \citep{Freytagetal2012}, \citet{DravinsLudwig2023} explored the temporal variability of granulation-driven RV signals in atomic and molecular lines. Achieving this requires spectral synthesis using a long time series of granulation covering a sufficient number of characteristic granular lifetimes. This task is, however, computationally demanding and is subject to effects from oscillations in the simulation domain, which covers only a small region on the star. These oscillations introduce an RV component of their own in the spectral lines which has to be effectively removed to extract the granulation-driven line changes \citep[see e.g.,][]{DravinsLudwig2023,Frameetal2025}. To circumvent this limitation, \citet{Sowmya25} introduced an efficient framework for predicting spectral line sensitivities by exploiting the ergodic nature of granulation, which allows time variability to be determined directly from the spatial variability of line profiles prevalent in a single snapshot of granulation. This framework was subsequently applied by \citet{Valera26} to stars cooler than the Sun, demonstrating how changes in granulation associated with stellar effective temperature modify spectral line sensitivities.

In this study, we apply the framework from \citet{Sowmya25} to investigate how stellar metallicity ($\rm M/H$) influences the sensitivity of spectral lines to granulation. The effect of the metallicity is twofold. Firstly, it directly affects the chemical composition, which alters strengths and formation heights of spectral lines. Secondly, it has an indirect effect on the stratification of the physical parameters, such as temperature, pressure, and density, by altering the opacity of radiation. Further, it also impacts convective flows by changing the efficiency of radiative energy transport. Together, all these effects are expected to modify the line sensitivity to granulation.

Here we explore how the line sensitivity to granulation depends on metallicity, and its implications for stellar RV mitigation strategies. The 3D MHD MURaM simulations used in this study and the subsequent spectral synthesis using the MPS-ATLAS radiative transfer code \citep{Witzke21} are described in Section~\ref{sec:modeling}. In the same section, we introduce the metrics we use to estimate the line sensitivity to granulation. In Section~\ref{sec:results} we present our results and discuss limitations of our approach. The conclusions are presented in Section~\ref{sec:conclusions}.

\section{Atmospheric models, spectral synthesis and sensitivity metrics}
\label{sec:modeling}

\begin{figure*}
    \centering
    \includegraphics[width=\textwidth]{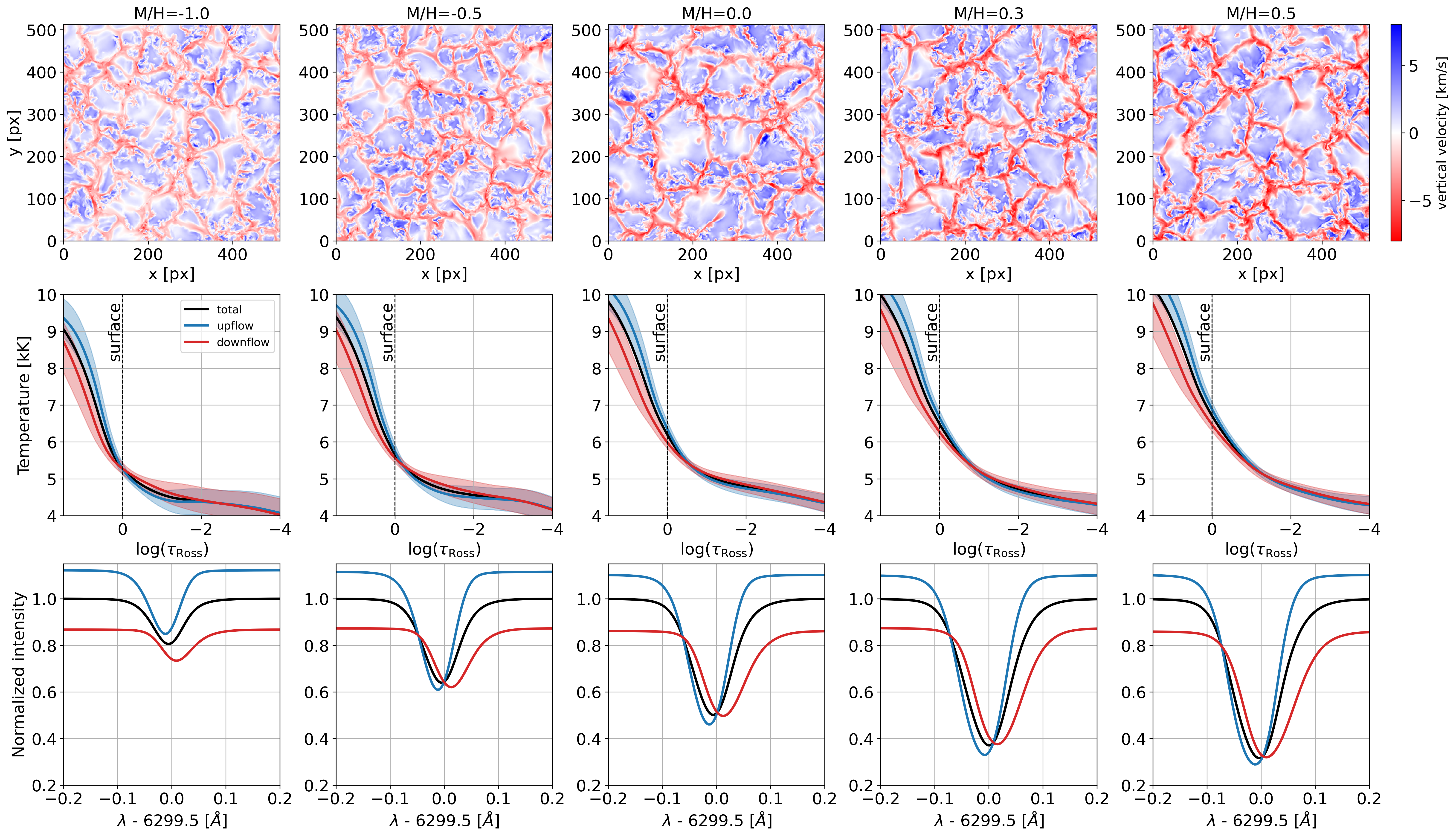}
    \caption{\textit{Top row:} The vertical velocity at the $\tau_\mathrm{Ross}=1$ layer for each metallicity. The color bar is capped at $\pm8$\,km/s. \textit{Middle row:} Spatially averaged temperature stratification (black line). The blue and red lines show spatially averaged temperature stratification for pixels corresponding to up- and down-flows based on the maps in the first row. Shaded regions depict the variation in the temperature at each depth. \textit{Bottom row:} The black lines show spectral profiles averaged over the whole spatial maps for a spectral line with $\log(gf)=-3$ and $E_{\rm low}=2.5$\,eV. The blue and red lines show the decomposition of the black line profiles into average profiles over the upflow and downflow regions, respectively. These profiles have been normalized to the mean continuum intensity of the whole map.}
    \label{fig:vz_avgT_profiles}
\end{figure*}

To investigate the dependence of spectral line sensitivity to granulation on metallicity, we consider atmospheric models of solar effective temperature and surface gravity with metallicity values $\rm M/H = \{-1.0,-0.5,0.0,0.3,0.5\}$. This chosen $\rm M/H$ range spans currently detected exoplanet host stars \citep{deLaverny25}. We refer to $\mathrm{M/H}=\{-1.0,-0.5\}$ as metal-poor and $\mathrm{M/H}=\{0.3,0.5\}$ as metal-rich models. We adapt 3D models of stellar granulation including a small-scale dynamo (SSD) for $\rm M/H = \{-1.0, 0.0, 0.5\}$ from \citet{Witzke23} and extend them to the intermediate cases $\rm M/H = \{-0.5, 0.3\}$ following the same setup as in \citet{Witzke23}. All these 3D models were simulated using the MURaM code \citep{Voegleretal2005,Rempel14}, following the box-in-a-star approach, which represents a small region on the star covering the upper convection zone and photosphere. The simulation domain across all $\rm M/H$ values has a horizontal extent of 9\,Mm x 9\,Mm with a resolution of 17.5\,km in both directions, resulting in a total of $512\times512$ surface elements (pixels). The vertical extent is 5\,Mm (from $\sim 4$\,Mm below the optical surface at $\tau_{500}=1$\footnote{Optical depth is unitless parameter describing the transparency of the medium. It is a product of a monochromatic opacity at a specific wavelength (in this case 500\,nm) and the physical path that light crosses.} to the temperature minimum) with a resolution of 10\,km. The simulation domain is chosen such that it contains at least a dozen granules in it (see, for example, top row of Figure~\ref{fig:vz_avgT_profiles}) for each $\rm M/H$. We refer to \citet{Witzke23} for the details on the simulation setup, the governing equations, and the boundary conditions.

Since stellar granulation is a stochastic phenomenon, its time-averaged properties are similar to the spatially averaged properties when evaluated on characteristic granulation time and length scales. This is the so-called ergodicity of granulation, which was demonstrated through a simple illustrative test in \citet{Sowmya25}. They showed that the spatial variability of line sensitivity to granulation, calculated from a single snapshot of granulation, is very similar to the temporal variability across different snapshots. Exploiting this property, we limit our analysis to the spatial variability of spectral lines across the granulation pattern, using a single MURaM snapshot from each of the five $\rm M/H$ cases.

We then post-process the MURaM snapshots to produce atmospheres for different viewing angles $\theta$, needed for spectral synthesis at different positions on the disk ranging from the disk center ($\mu=\cos(\theta)=1.0$) to the limb ($\mu=0.1$), with a step size of 0.1 in $\mu$. For this, we rotate the atmosphere about a pivot axis anchored at the mean optical surface to the required $\theta$. Then we compute the temperature, pressure, density, and velocity along the line of sight of the rotated atmosphere represented on the equal Rosseland mean optical depth grid.

The spatial distribution of the vertical velocity at the $\tau_\mathrm{Ross}=1$\footnote{Throughout the text we would use the mean Rosseland optical depth $\tau_\mathrm{Ross}$ to describe the atmospheric stratification. The $\tau_\mathrm{Ross}$ is computed using the mean Rosseland opacity.} layer at $\mu=1.0$ for each $\rm M/H$ is displayed in the top row of Figure~\ref{fig:vz_avgT_profiles}. The velocity maps clearly show the increase in the convective velocities with increasing $\rm M/H$. Using these velocity maps, we classify pixels into two groups corresponding to the intergranular lanes (down-flows) and granules (up-flows). The variation of the mean temperature within intergranular lanes and granules, along with the mean from the whole atmosphere, are shown as a function of $\log\tau_\mathrm{Ross}$ in the middle row of Figure~\ref{fig:vz_avgT_profiles} for the five $\rm M/H$ cases. We note an increase in the spatially averaged temperature and a decrease in the temperature gradient with increasing $\rm M/H$.

We synthesize emergent spectra assuming local thermodynamic equilibrium (LTE) with the MPS-ATLAS code \citep{Witzke21} using the post-processed atmospheric models. The combination of the MURaM atmospheric models and spectra from the MPS-ATLAS code has been used in many studies \citep[e.g.][]{Frameetal2025,Sowmya25,Valera26,Florian26,Frame26} and extensively tested in \citet{Witzke24}, highlighting their suitability for the present work. 

Owing to the action of the small-scale dynamo, the magnetic field strength at the surface of the atmospheric models is less than 100\,G \citep{Witzke23}. This magnetic field can produce a small line broadening due to the Zeeman effect. This effect splits spectral lines into multiple components symmetrically around the line's central wavelength, assuming that all the components are formed at the same height in the atmosphere. For the simplicity of the radiative transfer calculations and owing to the weak magnetic field in the atmospheric models, we opted to neglect the Zeeman broadening of the line profiles.

For the exploration of the line sensitivity, we resort to line transitions occurring in the neutral and ionized artificial sowmium element introduced in \citet{Sowmya25}, with the central wavelength at $\lambda_{\rm vacuum}=629.953$\,nm. The sowmium element is identical in its characteristics to the iron element, such that they have the same elemental abundance for solar metallicity, ionization potentials, atomic number, and weight. The main difference between iron and sowmium elements is in the line transitions. We constructed a set of sowmium line transitions that are characterized by the transition probability, $\log(gf)$, and the lower level excitation potential, $E_{\rm low}$. The transition probability of neutral and ionized sowmium lines is in range from $-6$ to $0$, where \elow{} for neutral lines is in the range $0$\,eV to $5$\,eV, and for ionized lines in the ragen $2$\,eV to $6$\,eV. Both \loggf{} and \elow{} are sampled in steps of $0.5$. The line broadening parameters used are the same as those for the neutral iron line at $\lambda_{\rm vacuum}=629.953$\,nm listed in the VALD3 database \citep{VALD,Ryabchikova2015}.

We precomputed the line opacities for each $\rm M/H$ as look-up tables with the spectral resolution of $R=2\,000\,000$. The look-up tables cover a broad range of temperature and pressure values typical of the conditions in the stellar upper convection zones and photospheres considered in this study. Using these look-up tables, we interpolate the opacities to the temperature and pressure at each point in the atmospheric model to solve the radiative transfer equation and obtain the emerging line profile. The so-computed sample of artificial lines represents a variety of spectral lines whose formation spans a broad range of heights in the photosphere and enables us to study the effects of granulation on the line shifts, strengths, and asymmetries. Example mean profiles for a line from the sample, with $\log(gf)=-3$ and $E_{\rm low}=2.5$\,eV, are shown in the bottom row of Figure~\ref{fig:vz_avgT_profiles} individually for the intergranular lanes, granules, and the whole atmosphere. 

The primary effect of $\rm M/H$ on the synthesized line profiles is a change in the line equivalent width. Increasing the metallicity increases the abundance of atoms contributing to the line opacity, thereby enhancing the opacity and reducing the emergent line intensity. The principal difference between the granular and intergranular line profiles is their relative depth. The hotter granules exhibit higher continuum intensities than the cooler intergranular lanes. Moreover, the shallower temperature gradients in the intergranular lanes result in, on average, shallower line profiles than those produced in the granules.

For the range of atomic parameters considered here, the spectral synthesis produces a wide variety of line profiles, including lines that are either very weak or very strong. Both of these types of lines will affect the line sensitivity metric because they are outliers. To exclude such outliers, we applied a selection criterion based on the central line depth of the spatially averaged line profile. We analyzed only spectral lines with the central depth between 0.1 and 0.7, for two reasons. First, weak lines have a low signal-to-noise ratio and are therefore particularly susceptible to observational uncertainties, which can contaminate the extraction of the RVs \citep[e.g.][]{Bouchyetal2001}. Although our synthetic spectra have effectively infinite signal-to-noise ratios, imposing a lower threshold on the line depth makes the analysis more representative of real observational conditions. Second, very strong lines are increasingly affected by departures from LTE, which are not accounted for in the MPS-ATLAS code. The non-LTE effects are most pronounced in strong lines that form high in the atmosphere above the solar granulation, where the atmospheric dynamics is governed primarily by the magnetic field rather than convective plasma motions. Modeling departures from LTE is beyond the scope of the present study. In contrast, the majority of the spectral lines used in stellar spectroscopy are of intermediate strength, for which LTE is a reasonable approximation. In addition, we required that the same line needs to satisfy the central depth criterion simultaneously in both the given $\rm M/H$ and the solar $\rm M/H=0.0$ case, independently at each viewing angle.

We then use the same methodology as in \citet{Sowmya25,Valera26} to quantify line sensitivity to granulation. Here we provide a brief description
for completeness and to facilitate the discussion of the results. 

Starting with the $512\times512$ emergent intensity spectra for a given spectral line, we average every consecutive $16\times16$ pixels to form a total of $32\times32$ spaxels. The spatial averaging of line profiles into coarser spaxels smears out abrupt changes in the line profiles caused by strong gradients in physical parameters, hence removing any outliers in the sensitivity measures. The resulting 1024 spaxels for each $\rm M/H$ contain a wide variety of line profiles arising from different mixtures of constituent structures such as intergranular lanes, granules, and bright points (which occasionally appear in the intergranular lanes due to the action of SSD). 

For each spaxel, we compute the line center of gravity wavelength, $\lambda_\mathrm{COG}$, and line equivalent width, $W$. Then, we calculated the metric to quantify line sensitivity to granulation, and opted to use the weighted standard deviation:
\begin{equation}
    \sigma^2_q = \frac{\sum_i^{N_s} w_i\cdot\left(q_i - \left<q\right>\right)^2}{(N_s-1)\sum_i^{N_s} w_i / N_s},
\label{eq:metric}
\end{equation}
where $\left<q\right>$ is:
\begin{equation}
    \left<q\right> = \frac{\sum_i^{N_s} w_i\cdot q_i}{\sum_i w_i}.
\end{equation}

Here $q_i$ denotes either the center of gravity or the equivalent width of spaxel $i$, and $N_s$ is the number of spaxels. The weighting coefficients, $w_i$, are proportional to the continuum intensity of each spaxel's spectrum. Because the spatial variation of the continuum intensity is assumed to be statistically equivalent to its temporal variation (i.e., under the ergodicity assumption for granulation), this weighting scheme is intended to reproduce the temporal fluctuations of the measured continuum flux across different observing nights.

We subsequently repeat this procedure for all the lines in the sample and for all 10 viewing angles. As the solar spectrum serves as a reference for any stellar studies, we opted to contrast our results with those obtained for $\rm M/H=0.0$, corresponding to the solar atmospheric model. Both $\sigma_\lambda$ and $\sigma_W$ are in wavelength units.

\section{Results}
\label{sec:results}

\subsection{Sowmium lines}

\begin{figure}
    \centering
    \includegraphics[width=\linewidth]{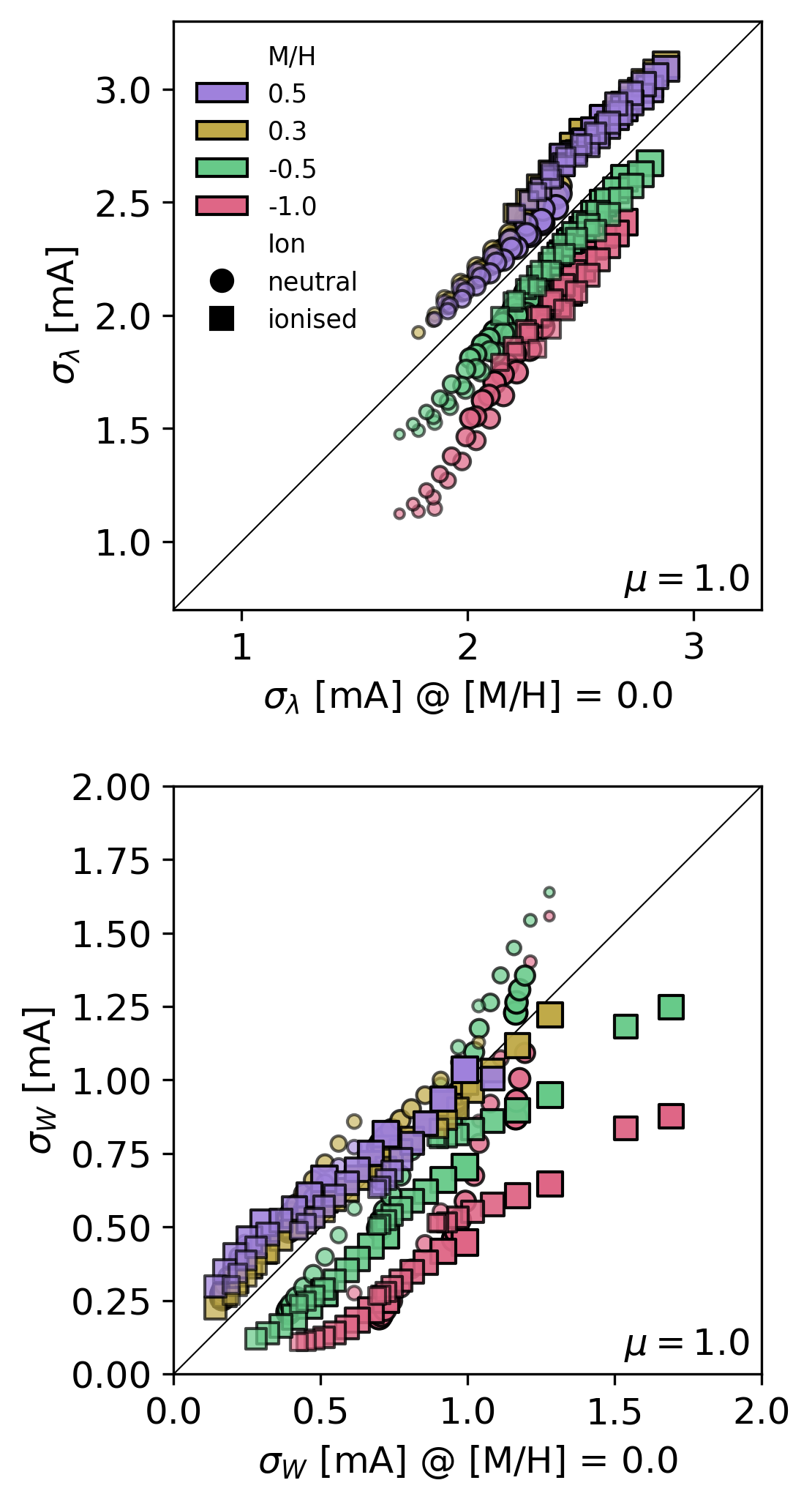}
    \caption{Scatter plots of the sensitivity metrics with metallicity at $\mu=1$ for the sowmium lines (fictional iron lines uniformly sampled in transition probability and lower excitation potential). The results from the solar $\mathrm{M/H}=0.0$ case are shown on the $x$-axis, while those from the non-solar metallicity cases are on the $y$-axis. \textit{Top panel:} the weighted standard deviation of the line center of gravity wavelength, $\sigma_\lambda$. \textit{Bottom panel:} the weighted standard deviation of the line equivalent width $\sigma_W$. The larger point size is related to the higher \elow{}, while transparent points correspond to the low \loggf{}. The diagonal black line represents a 1-1 comparison. Only lines that have line core depth between $0.1$ and $0.7$ in the solar and corresponding metallicity are shown.}
    \label{fig:proxy_comparison}
\end{figure}

The top panel of Figure\,\ref{fig:proxy_comparison} compares $\sigma_\lambda$ for the solar ($\mathrm{M/H}=0.0$) and non-solar metallicity cases at $\mu=1$. Spectral lines computed from the metal-rich models exhibit larger values of $\sigma_\lambda$ than their solar counterparts systematically, whereas those formed in the metal-poor models show smaller values. This trend is observed for both neutral and ionized lines. The ionized lines consistently display larger $\sigma_\lambda$ than the neutral lines, in agreement with the findings of \citet{Sowmya25} for the solar case. This is the consequence of stronger convective flows at deeper layers of the atmosphere where ionized lines are predominantly formed for the solar effective temperature. In contrast, the neutral lines form at higher layers in the atmosphere where the convective motion is weaker, causing them to exhibit smaller $\sigma_\lambda$.

The small separation between the $M/H=0.3$ and $M/H=0.5$ cases reflects their negligible differences in the stratification of the physical parameters (see Figure\,\ref{fig:avgT_RMSvz} in Appendix\,\ref{sec:app_atmos}). In addition, we note that the separation between neutral and ionized lines in the metal-rich cases indicates that they form at distinct atmospheric layers compared to the metal-poor models. The metal-rich models exhibit steeper temperature gradients above the surface, allowing photons to escape from shallower layers compared to the metal-poor models.

A similar metallicity dependence is seen for the equivalent-width sensitivity metric, $\sigma_W$, shown in the bottom panel of Figure\,\ref{fig:proxy_comparison}, although in this case no clear separation between neutral and ionized lines is observed. The principal effect of metallicity is therefore to rescale the amplitudes of both sensitivity metrics, while preserving the relative behavior of neutral and ionized spectral lines.

\begin{figure}
    \centering
    \includegraphics[width=\linewidth]{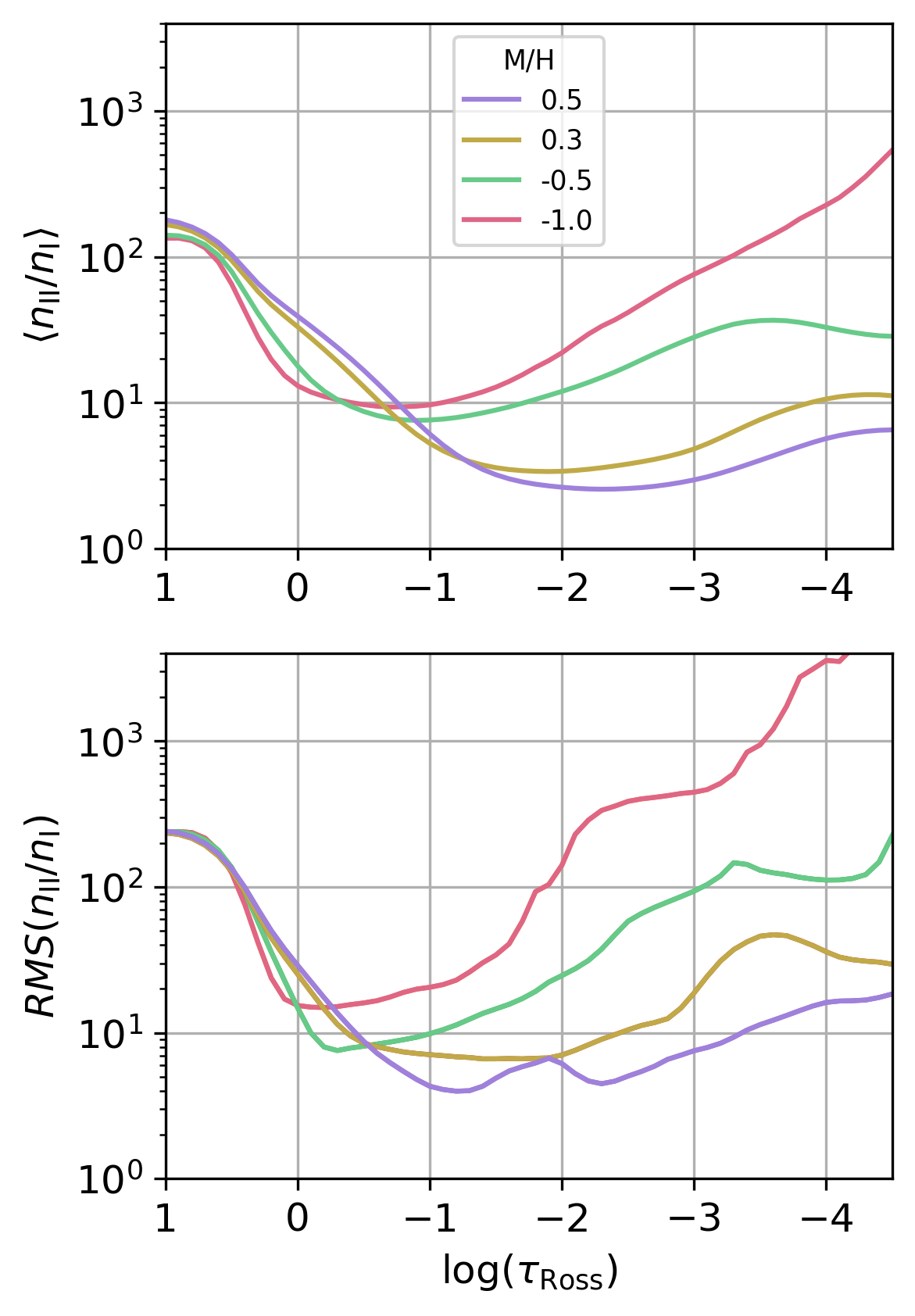}
    \caption{The ratio between number densities of ionized and neutral sowmium atoms, $n_\mathrm{II}/n_\mathrm{I}$, for all metallicity cases. \textit{Top panel:} Variation of the spatially averaged values with $\log(\tau_\mathrm{Ross})$. \textit{Bottom panel:} Variation of the RMS values with $\log(\tau_\mathrm{Ross})$. The color coding is the same as in Figure\,\ref{fig:proxy_comparison}.}
    \label{fig:ion_fraction}
\end{figure}

The persistence of the neutral versus ionized ordering across metallicities can be understood from the ionization balance of sowmium. Figure\,\ref{fig:ion_fraction} shows the ratio between the number densities of ionized and neutral sowmium atoms, $n_\mathrm{II}/n_\mathrm{I}$, together with its RMS variation for all non-solar metallicity models. The ratio remains well above unity at all heights for all non-solar $\mathrm{M/H}$ models as shown, indicating that sowmium is predominantly found in its first ionization stage for the solar effective temperature, irrespective of metallicity. The increase in the number density of ionized sowmium atoms is driven by the decrease in gas density with height, which reduces the electron number density and shifts the ionization balance toward the ionized state.

Neutral sowmium remains the minority species, preserving the relative response of the two ionization stages to granulation. The same argument is expected to apply to other low first-ionization-potential elements commonly observed in near-solar effective temperature stellar spectra, such as titanium and vanadium. Thus, while metallicity changes the amplitude of the granulation signatures, it does not modify the underlying physical mechanism responsible for the different behavior of neutral and ionized lines.

We remark that for simplicity, we disregard non-LTE effects. For example, neutral iron in the solar atmosphere is subject to UV overionization, which reduces the Fe\,I population and weakens Fe\,I lines \citep[e.g.][]{Shchukina01}. Such effects would make the neutral species an even smaller minority, reinforcing the dominance of the ionized stage shown in Figure\,\ref{fig:ion_fraction}. These effects are therefore not expected to change the relative sensitivity of neutral and ionized lines to granulation.

We next explain why the amplitudes of the sensitivity metrics increase systematically with metallicity. We first consider the origin of the increase in $\sigma_\lambda$. Previous studies based on the HD and MHD simulations have shown that increasing metallicity leads to stronger convective velocities because the higher opacity reduces the efficiency of radiative energy transport. This modifies the atmospheric structure and shifts the photosphere to lower density layers, where larger convective velocities are required to transport the stellar flux \citep[e.g.][]{Ludwigetal2009,Magic2013,Witzke23}. Consequently, the RMS vertical velocity increases with metallicity \citep{Witzke23}. Since $\sigma_\lambda$ is a measure of the line shifts induced by granulation, the increase in RMS vertical velocity with metallicity leads to larger amplitudes of $\sigma_\lambda$.

\begin{figure}
    \centering
    \includegraphics[width=\linewidth]{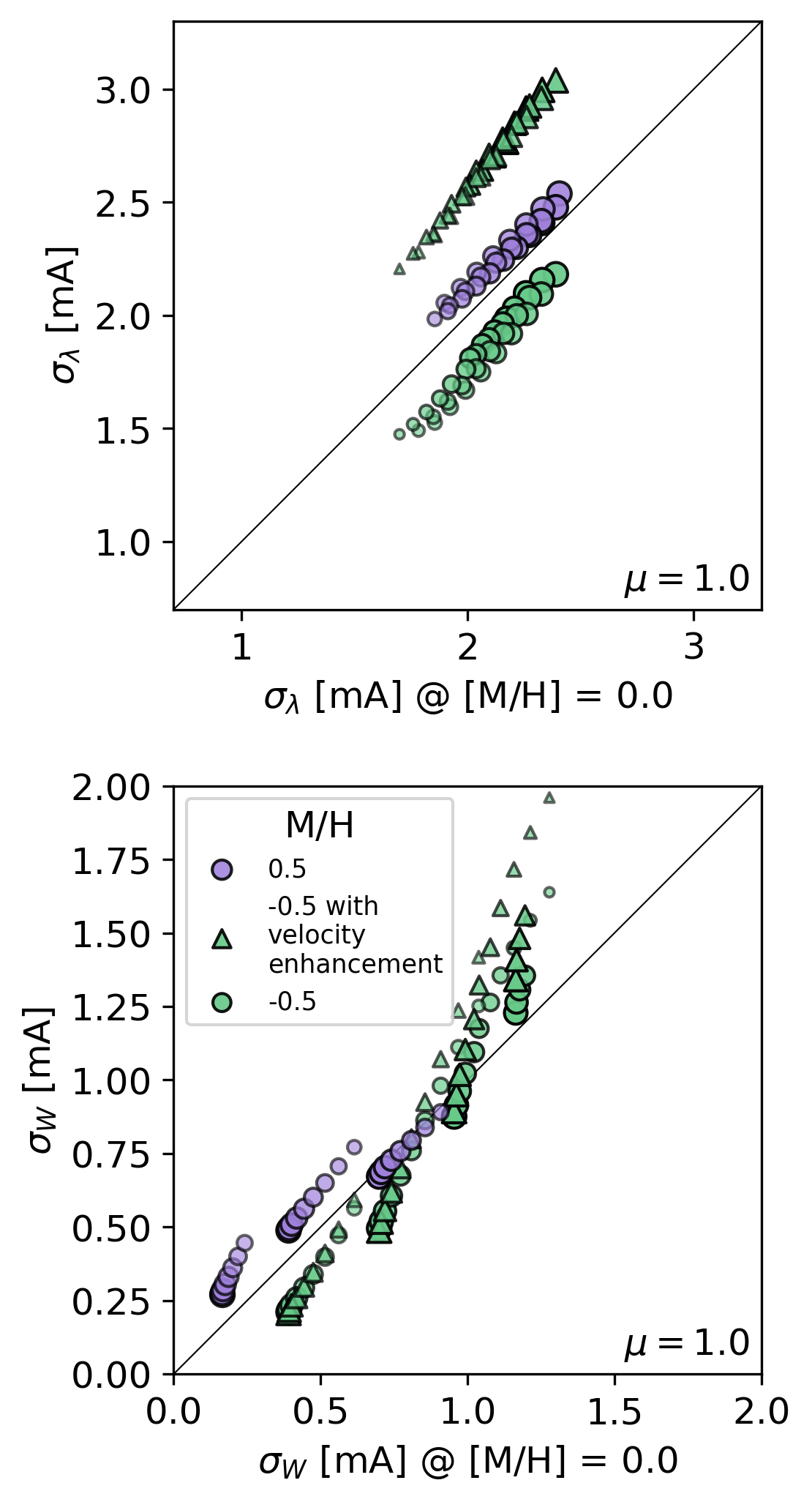}
    \caption{Same as Fig.\,\ref{fig:proxy_comparison} but comparing the results from the $M/H=0.5$ (purple circles) and the velocity-enhanced $M/H=-0.5$ model (green triangles). For the comparison, we also show the results from the nominal $M/H=-0.5$ model in green circles. \textit{Top panel:} The comparison for $\sigma_\lambda$. \textit{Bottom panel:} The comparison for $\sigma_W$. Size and the opacity of the points have the same meaning as in Fig.\,\ref{fig:proxy_comparison}.}
    \label{fig:los_enchancement_test}
\end{figure}

To investigate the role of the velocity field further, we conducted the following experiment. We computed the ratio of the RMS vertical velocity between the $\mathrm{M/H}=0.5$ and $\mathrm{M/H}=-0.5$ models at every depth point in the atmosphere. We then multiplied the vertical velocity in the $\mathrm{M/H}=-0.5$ model by this ratio, thereby enhancing the velocity field while leaving the thermal structure unchanged. We subsequently recomputed all neutral spectral lines and their sensitivity metrics for the velocity-enhanced $\mathrm{M/H}=-0.5$ model. As shown in Figure\,\ref{fig:los_enchancement_test}, increasing the vertical velocity leads to a substantial increase in $\sigma_\lambda$, demonstrating that this metric is primarily controlled by the strength of convective motions. Larger convective velocities produce steeper velocity gradients and greater asymmetries in the spaxel line profiles, resulting in larger shifts of the line COG wavelength. We note that the resulting $\sigma_\lambda$ values are even larger than those of the original $\mathrm{M/H}=0.5$ model. We attribute this to the lower temperature RMS of the metal-rich atmosphere (see middle row of Figure\,\ref{fig:vz_avgT_profiles}), which produces intrinsically narrower spaxel line profiles and in turn reduces $\sigma_\lambda$.

The velocity-enhancement experiment also provides insight into the origin of the increase in $\sigma_W$ shown in the bottom panel of Figure\,\ref{fig:proxy_comparison}. In contrast to $\sigma_\lambda$, increasing the convective velocity field produces only a minor change in $\sigma_W$ (Figure\,\ref{fig:los_enchancement_test}), indicating that this metric is only weakly affected by the strength of convective motions. Instead, the increase in $\sigma_W$ with metallicity is primarily driven by changes in the atmospheric structure and line formation. Increasing the metallicity enhances the abundance of absorbing species and increases the atmospheric opacity, producing stronger and deeper spectral lines, as illustrated in Figure\,\ref{fig:vz_avgT_profiles}. At the same time, the increased opacity shifts the line formation region to higher atmospheric layers, where the thermodynamic conditions differ from those in the metal-poor models. Since equivalent-width variations are governed mainly by changes in the local thermodynamic state through their influence on the line opacity and source function, these modifications to the line formation process naturally lead to the larger $\sigma_W$ values found in the metal-rich models.

\subsection{Center-to-limb sensitivity variation}

\begin{figure}
    \centering
    \includegraphics[width=\linewidth]{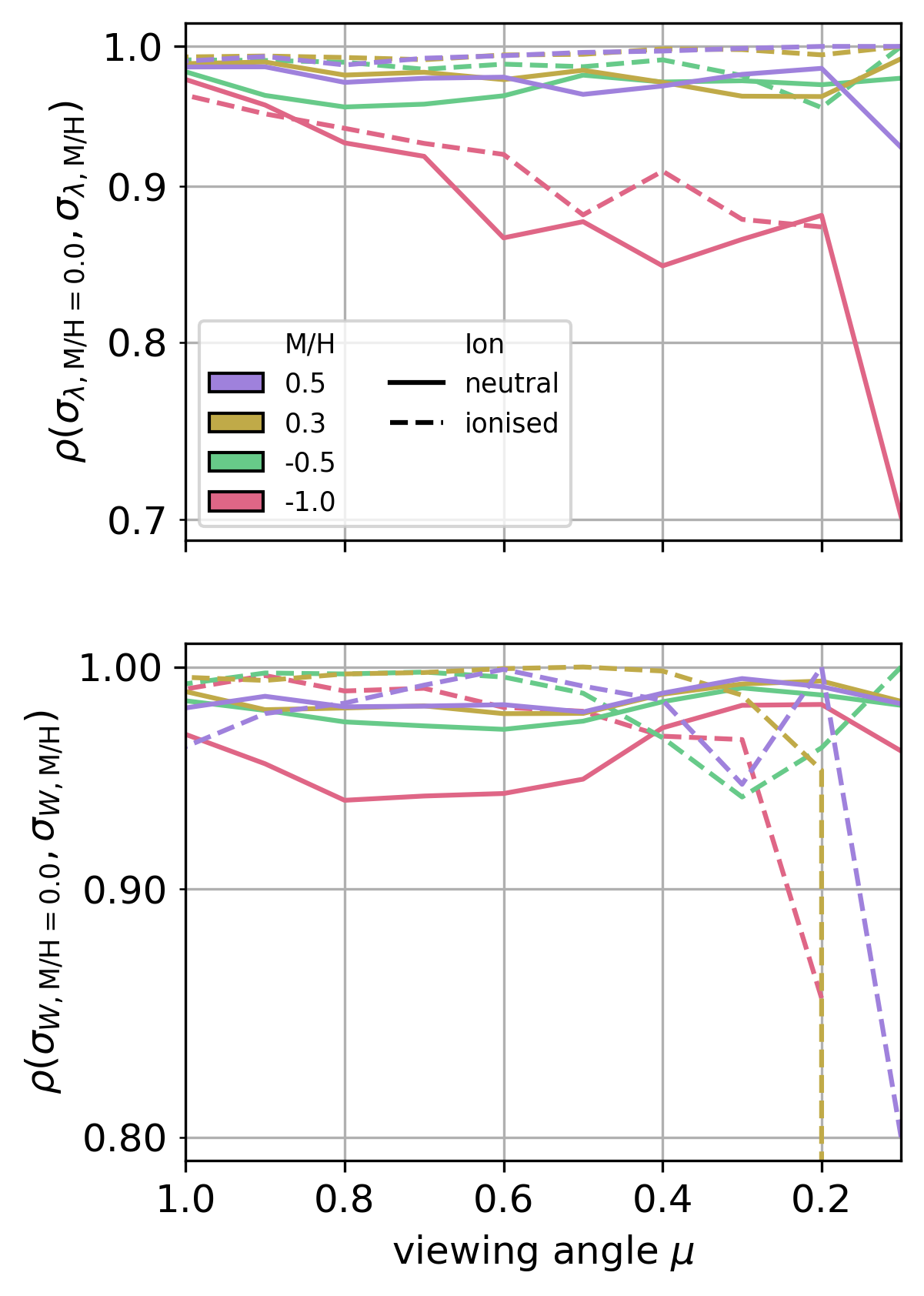}
    \caption{The Spearman rank correlation coefficient for $\sigma_\lambda$ (top panel) and $\sigma_W$ (bottom panel) for lines from the solar $\mathrm{M/H}=0.0$ and non-solar metallicity $\mathrm{M/H}$ models. The solid and dashed lines respectively show the correlation coefficient for the neutral and the ionized sowmium lines across all viewing angles.}
    \label{fig:spearman}
\end{figure}

Until now, we discussed the sensitivity of spectral lines to granulation at the disc center. Here, we examine whether the relative sensitivity of different spectral lines is preserved toward the stellar limb. To this end, we repeated the analysis presented in Figure\,\ref{fig:proxy_comparison} for all viewing angles. We then computed the Spearman rank correlation coefficient between the $\sigma_\lambda$ values of the neutral and ionized sowmium lines from the solar $\mathrm{M/H}=0.0$ model and the corresponding $\sigma_\lambda$ values from the non-solar metallicity models, as shown in Figure\,\ref{fig:spearman}.

The Spearman rank correlation coefficient remains high at all viewing angles for both $\sigma_\lambda$ and $\sigma_W$. The decrease in the correlation toward the limb arises because the line of sight intersects multiple upflow and downflow regions with substantially different temperature stratifications. These variations in the atmospheric structure produce line profiles that can deviate significantly from standard Voigt shapes and, in some cases, exhibit emission in the line core. We therefore excluded profiles with strong core emission, which resulted in an increase in the correlation coefficient. However, some profiles also exhibit emission in the line wings, although the intensity remains below the continuum level. This primarily affects the equivalent width and likely accounts for the larger decrease in the correlation for $\sigma_W$ compared with $\sigma_\lambda$. Ionized lines are more strongly affected by the atmospheric velocity structure closer to the limb than neutral lines because they form at greater depths, where they sample a larger number of upflowing and downflowing regions, resulting in more complex line profiles.

The strongest decrease in the correlation coefficient occurs at a metallicity of $M/H=-1.0$. At this metallicity, the vertical velocity exhibits a substantially different height dependence compared with the solar-metallicity case (see Figure\,\ref{fig:avgT_RMSvz} in Appendix\,\ref{sec:app_atmos}). This difference is most clearly reflected in the pronounced decrease in the $\sigma_\lambda$ correlation toward the limb.

\medskip

These results demonstrate that the sensitivity of spectral lines to granulation is largely preserved across atmospheric models with different $\mathrm{M/H}$ values and remains robust from the stellar disc center to the limb. In other words, lines which are sensitive to granulation in solar metallicity remain sensitive to granulation in non-solar metallicity models. Based on these findings, we conclude that the trends governing the sensitivity of spectral lines to granulation in the solar atmosphere can be reliably extended to atmospheres with non-solar metallicities, provided they have the same effective temperature and surface gravity as the Sun.

\section{Conclusions}
\label{sec:conclusions}

We investigated the effect of metallicity on the sensitivity of spectral lines to granulation using 3D magnetohydrodynamic small-scale dynamo models of the upper convection zone and photosphere with metallicities $\mathrm{M/H}={-1.0,-0.5,0.0,0.3,0.5}$ but solar effective temperature and surface gravity. Neutral and ionized sowmium lines were synthesized in local thermodynamic equilibrium for viewing angles ranging from the stellar disc center ($\mu=1$) to the limb ($\mu=0.1$). The sensitivity to granulation was quantified using the continuum-intensity-weighted standard deviations of the line center-of-gravity wavelength, $\sigma_\lambda$, and equivalent width, $\sigma_W$.

We showed that both $\sigma_\lambda$ and $\sigma_W$ increase systematically with metallicity while preserving the same overall behavior for neutral and ionized lines. The increase in $\sigma_\lambda$ is primarily driven by the stronger convective velocities in metal-rich atmospheres, whereas the increase in $\sigma_W$ results mainly from metallicity-dependent changes in line opacity. Thus, we can say that the metallicity indirectly increases the line sensitivity to granulation by altering the physical conditions in the stellar atmosphere (e.g., velocity is enhanced with metallicity). However, we further showed that the relative sensitivity of spectral lines to granulation is largely preserved across the stellar disc and over the full metallicity range considered. Consequently, the line sensitivity to granulation established for the solar metallicity can be reliably extended to atmospheres with non-solar metallicities, provided they have the same effective temperature and surface gravity as the Sun.

These findings support the use of solar-based granulation studies to interpret observations of stars with different metallicities, which is particularly timely in the context of high-precision radial-velocity surveys and the upcoming PLATO mission \citep{Rauer25}, which is expected to significantly increase the number of targets for RV follow-ups.

\begin{acknowledgments}
We acknowledge support from the European Research Council (ERC) under the European Union’s Horizon 2020 research and innovation program (grant No. 101118581). VV acknowledges support from the Max Planck Society under the grant ``PLATO Science'' and from the German Aerospace Center under ``PLATO Data Center'' grant 50OO1501. This work has made use of the VALD database, operated at Uppsala University, the Institute of Astronomy RAS in Moscow, and the University of Vienna. This study has made use of SAO/NASA Astrophysics Data System's bibliographic services. We gratefully acknowledge the computational resources provided by the Raven supercomputer systems of the Max Planck Computing and Data Facility (MPCDF) in Garching, Germany.
\end{acknowledgments}

\begin{contribution}

DV, SK, and VV conceptualized the project. VV produced the atmospheric models. DV produced the results and created the first manuscript draft. All authors contributed to manuscript writing.


\end{contribution}

%
\facilities{ We gratefully acknowledge the computational resources provided by the  Raven supercomputer systems of the Max Planck Computing and Data Facility (MPCDF) in Garching, Germany.
}

\software{
matplotlib \citep{matplotlib}, 
numpy \citep{numpy}, 
scipy \citep{scipy}
}


\bibliography{references}{}

\begin{thebibliography}{}
\expandafter\ifx\csname natexlab\endcsname\relax\def\natexlab#1{#1}\fi
\providecommand{\url}[1]{\href{#1}{#1}}
\providecommand{\dodoi}[1]{doi:~\href{http://doi.org/#1}{\nolinkurl{#1}}}
\providecommand{\doeprint}[1]{\href{http://ascl.net/#1}{\nolinkurl{http://ascl.net/#1}}}
\providecommand{\doarXiv}[1]{\href{https://arxiv.org/abs/#1}{\nolinkurl{https://arxiv.org/abs/#1}}}

\bibitem[{K. {Al Moulla} {et~al.}(2022){Al Moulla}, {Dumusque}, {Cretignier}, {Zhao}, \& {Valenti}}]{AlMoulla22}
{Al Moulla}, K., {Dumusque}, X., {Cretignier}, M., {Zhao}, Y., \& {Valenti}, J.~A. 2022, \bibinfo{title}{{Measuring precise radial velocities on individual spectral lines. III. Dependence of stellar activity signal on line formation temperature},} \aap, 664, A34, \dodoi{10.1051/0004-6361/202243276}

\bibitem[{G. {Anglada-Escud{\'e}} \& R.~P. {Butler}(2012){Anglada-Escud{\'e}} \& {Butler}}]{Anglada-Escude12}
{Anglada-Escud{\'e}}, G., \& {Butler}, R.~P. 2012, \bibinfo{title}{{The HARPS-TERRA Project. I. Description of the Algorithms, Performance, and New Measurements on a Few Remarkable Stars Observed by HARPS},} \apjs, 200, 15, \dodoi{10.1088/0067-0049/200/2/15}

\bibitem[{A. {Baranne} {et~al.}(1996){Baranne}, {Queloz}, {Mayor}, {Adrianzyk}, {Knispel}, {Kohler}, {Lacroix}, {Meunier}, {Rimbaud}, \& {Vin}}]{Baranne96}
{Baranne}, A., {Queloz}, D., {Mayor}, M., {et~al.} 1996, \bibinfo{title}{{ELODIE: A spectrograph for accurate radial velocity measurements.},} \aaps, 119, 373

\bibitem[{F. {Bouchy} {et~al.}(2001){Bouchy}, {Pepe}, \& {Queloz}}]{Bouchyetal2001}
{Bouchy}, F., {Pepe}, F., \& {Queloz}, D. 2001, \bibinfo{title}{{Fundamental photon noise limit to radial velocity measurements},} \aap, 374, 733, \dodoi{10.1051/0004-6361:20010730}

\bibitem[{H.~M. {Cegla} {et~al.}(2013){Cegla}, {Shelyag}, {Watson}, \& {Mathioudakis}}]{Ceglaetal2013}
{Cegla}, H.~M., {Shelyag}, S., {Watson}, C.~A., \& {Mathioudakis}, M. 2013, \bibinfo{title}{{Stellar Surface Magneto-convection as a Source of Astrophysical Noise. I. Multi-component Parameterization of Absorption Line Profiles},} \apj, 763, 95, \dodoi{10.1088/0004-637X/763/2/95}

\bibitem[{H.~M. {Cegla} {et~al.}(2019){Cegla}, {Watson}, {Shelyag}, {Mathioudakis}, \& {Moutari}}]{Ceglaetal2019}
{Cegla}, H.~M., {Watson}, C.~A., {Shelyag}, S., {Mathioudakis}, M., \& {Moutari}, S. 2019, \bibinfo{title}{{Stellar Surface Magnetoconvection as a Source of Astrophysical Noise. III. Sun-as-a-Star Simulations and Optimal Noise Diagnostics},} \apj, 879, 55, \dodoi{10.3847/1538-4357/ab16d3}

\bibitem[{H.~M. {Cegla} {et~al.}(2018){Cegla}, {Watson}, {Shelyag}, {Chaplin}, {Davies}, {Mathioudakis}, {Palumbo}, {Saar}, \& {Haywood}}]{Ceglaetal2018}
{Cegla}, H.~M., {Watson}, C.~A., {Shelyag}, S., {et~al.} 2018, \bibinfo{title}{{Stellar Surface Magneto-convection as a Source of Astrophysical Noise. II. Center-to-limb Parameterization of Absorption Line Profiles and Comparison to Observations},} \apj, 866, 55, \dodoi{10.3847/1538-4357/aaddfc}

\bibitem[{A. {Collier Cameron} {et~al.}(2021){Collier Cameron}, {Ford}, {Shahaf}, {Aigrain}, {Dumusque}, {Haywood}, {Mortier}, {Phillips}, {Buchhave}, {Cecconi}, {Cegla}, {Cosentino}, {Cr{\'e}tignier}, {Ghedina}, {Gonz{\'a}lez}, {Latham}, {Lodi}, {L{\'o}pez-Morales}, {Micela}, {Molinari}, {Pepe}, {Piotto}, {Poretti}, {Queloz}, {Juan}, {S{\'e}gransan}, {Sozzetti}, {Szentgyorgyi}, {Thompson}, {Udry}, \& {Watson}}]{CollierCameron21}
{Collier Cameron}, A., {Ford}, E.~B., {Shahaf}, S., {et~al.} 2021, \bibinfo{title}{{Separating planetary reflex Doppler shifts from stellar variability in the wavelength domain},} \mnras, 505, 1699, \dodoi{10.1093/mnras/stab1323}

\bibitem[{J. {Crass} {et~al.}(2021){Crass}, {Gaudi}, {Leifer}, {Beichman}, {Bender}, {Blackwood}, {Burt}, {Callas}, {Cegla}, {Diddams}, {Dumusque}, {Eastman}, {Ford}, {Fulton}, {Gibson}, {Halverson}, {Haywood}, {Hearty}, {Howard}, {Latham}, {L{\"o}hner-B{\"o}ttcher}, {Mamajek}, {Mortier}, {Newman}, {Plavchan}, {Quirrenbach}, {Reiners}, {Robertson}, {Roy}, {Schwab}, {Seifahrt}, {Szentgyorgyi}, {Terrien}, {Teske}, {Thompson}, \& {Vasisht}}]{Crassetal2021}
{Crass}, J., {Gaudi}, B.~S., {Leifer}, S., {et~al.} 2021, \bibinfo{title}{{Extreme Precision Radial Velocity Working Group Final Report},} arXiv e-prints, arXiv:2107.14291, \dodoi{10.48550/arXiv.2107.14291}

\bibitem[{M. {Cretignier} {et~al.}(2020){Cretignier}, {Dumusque}, {Allart}, {Pepe}, \& {Lovis}}]{Cretignier20}
{Cretignier}, M., {Dumusque}, X., {Allart}, R., {Pepe}, F., \& {Lovis}, C. 2020, \bibinfo{title}{{Measuring precise radial velocities on individual spectral lines. II. Dependance of stellar activity signal on line depth},} \aap, 633, A76, \dodoi{10.1051/0004-6361/201936548}

\bibitem[{P. {de Laverny} {et~al.}(2025){de Laverny}, {Ligi}, {Crida}, {Recio-Blanco}, \& {Palicio}}]{deLaverny25}
{de Laverny}, P., {Ligi}, R., {Crida}, A., {Recio-Blanco}, A., \& {Palicio}, P.~A. 2025, \bibinfo{title}{{The Gaia spectroscopic catalogue of exoplanets and host stars},} \aap, 699, A100, \dodoi{10.1051/0004-6361/202554739}

\bibitem[{D. {Dravins}(1982){Dravins}}]{Dravins1982}
{Dravins}, D. 1982, \bibinfo{title}{{Photospheric spectrum line asymmetries and wavelength shifts},} \araa, 20, 61, \dodoi{10.1146/annurev.aa.20.090182.000425}

\bibitem[{D. {Dravins}(1987{\natexlab{a}}){Dravins}}]{Dravins1987a}
{Dravins}, D. 1987{\natexlab{a}}, \bibinfo{title}{{Stellar granulation. I - The observability of stellar photospheric convection},} \aap, 172, 200

\bibitem[{D. {Dravins}(1987{\natexlab{b}}){Dravins}}]{Dravins1987b}
{Dravins}, D. 1987{\natexlab{b}}, \bibinfo{title}{{Stellar granulation. II. Stellar photospheric line asymmetries.},} \aap, 172, 211

\bibitem[{D. {Dravins} {et~al.}(1981){Dravins}, {Lindegren}, \& {Nordlund}}]{Dravinsetal1981}
{Dravins}, D., {Lindegren}, L., \& {Nordlund}, A. 1981, \bibinfo{title}{{Solar granulation - Influence of convection on spectral line asymmetries and wavelength shifts},} \aap, 96, 345

\bibitem[{D. {Dravins} \& H.-G. {Ludwig}(2023){Dravins} \& {Ludwig}}]{DravinsLudwig2023}
{Dravins}, D., \& {Ludwig}, H.-G. 2023, \bibinfo{title}{{Solar photospheric spectrum microvariability. I. Theoretical searches for proxies of radial-velocity jittering},} \aap, 679, A3, \dodoi{10.1051/0004-6361/202347142}

\bibitem[{X. {Dumusque}(2018){Dumusque}}]{Dumusque18}
{Dumusque}, X. 2018, \bibinfo{title}{{Measuring precise radial velocities on individual spectral lines. I. Validation of the method and application to mitigate stellar activity},} \aap, 620, A47, \dodoi{10.1051/0004-6361/201833795}

\bibitem[{D.~A. {Fischer} {et~al.}(2016){Fischer}, {Anglada-Escude}, {Arriagada}, {Baluev}, {Bean}, {Bouchy}, {Buchhave}, {Carroll}, {Chakraborty}, {Crepp}, {Dawson}, {Diddams}, {Dumusque}, {Eastman}, {Endl}, {Figueira}, {Ford}, {Foreman-Mackey}, {Fournier}, {F{\H{u}}r{\'e}sz}, {Gaudi}, {Gregory}, {Grundahl}, {Hatzes}, {H{\'e}brard}, {Herrero}, {Hogg}, {Howard}, {Johnson}, {Jorden}, {Jurgenson}, {Latham}, {Laughlin}, {Loredo}, {Lovis}, {Mahadevan}, {McCracken}, {Pepe}, {Perez}, {Phillips}, {Plavchan}, {Prato}, {Quirrenbach}, {Reiners}, {Robertson}, {Santos}, {Sawyer}, {Segransan}, {Sozzetti}, {Steinmetz}, {Szentgyorgyi}, {Udry}, {Valenti}, {Wang}, {Wittenmyer}, \& {Wright}}]{Fisher16}
{Fischer}, D.~A., {Anglada-Escude}, G., {Arriagada}, P., {et~al.} 2016, \bibinfo{title}{{State of the Field: Extreme Precision Radial Velocities},} \pasp, 128, 066001, \dodoi{10.1088/1538-3873/128/964/066001}

\bibitem[{G. {Frame} {et~al.}(2025){Frame}, {Cegla}, {Witzke}, {Lagae}, {Palumbo}, {Shelyag}, {Watson}, \& {Shapiro}}]{Frameetal2025}
{Frame}, G., {Cegla}, H.~M., {Witzke}, V., {et~al.} 2025, \bibinfo{title}{{Towards understanding stellar variability at the sub m/s level: isolating granulation signals in synthetic spectral lines},} \mnras, 539, 2248, \dodoi{10.1093/mnras/staf639}

\bibitem[{G. {Frame} {et~al.}(2026){Frame}, {Cegla}, {Lagae}, {Witzke}, {Watson}, {Shelyag}, {Panwar}, {Palumbo}, \& {Shapiro}}]{Frame26}
{Frame}, G., {Cegla}, H.~M., {Lagae}, C., {et~al.} 2026, \bibinfo{title}{{Synthetic disc-integrated absorption lines isolating stellar granulation for high-precision RV studies},} \mnras, 547, stag418, \dodoi{10.1093/mnras/stag418}

\bibitem[{B. {Freytag} {et~al.}(2012){Freytag}, {Steffen}, {Ludwig}, {Wedemeyer-B{\"o}hm}, {Schaffenberger}, \& {Steiner}}]{Freytagetal2012}
{Freytag}, B., {Steffen}, M., {Ludwig}, H.~G., {et~al.} 2012, \bibinfo{title}{{Simulations of stellar convection with CO5BOLD},} Journal of Computational Physics, 231, 919, \dodoi{10.1016/j.jcp.2011.09.026}

\bibitem[{D.~F. {Gray}(1981){Gray}}]{Gray1981}
{Gray}, D.~F. 1981, \bibinfo{title}{{Asymmetries in the spectral lines of Procyon},} \apj, 251, 583, \dodoi{10.1086/159501}

\bibitem[{D.~F. {Gray}(1982){Gray}}]{Gray1982}
{Gray}, D.~F. 1982, \bibinfo{title}{{Observations of spectral line asymmetries and convective velocities in F, G and K stars.},} \apj, 255, 200, \dodoi{10.1086/159818}

\bibitem[{D.~F. {Gray}(1983){Gray}}]{Gray1983}
{Gray}, D.~F. 1983, \bibinfo{title}{{On the constancy of spectral-line bisectors.},} \pasp, 95, 252, \dodoi{10.1086/131152}

\bibitem[{D.~F. {Gray}(1989){Gray}}]{Gray1989}
{Gray}, D.~F. 1989, \bibinfo{title}{{The Morphology of Reversed Spectral Line Bisectors},} \pasp, 101, 832, \dodoi{10.1086/132503}

\bibitem[{J.~D. {Hunter}(2007){Hunter}}]{matplotlib}
{Hunter}, J.~D. 2007, \bibinfo{title}{{Matplotlib: A 2D Graphics Environment},} Computing in Science and Engineering, 9, 90, \dodoi{10.1109/MCSE.2007.55}

\bibitem[{C. {Jurgenson} {et~al.}(2016){Jurgenson}, {Fischer}, {McCracken}, {Sawyer}, {Szymkowiak}, {Davis}, {Muller}, \& {Santoro}}]{Jurgensonetal2016}
{Jurgenson}, C., {Fischer}, D., {McCracken}, T., {et~al.} 2016, \bibinfo{title}{{EXPRES: a next generation RV spectrograph in the search for earth-like worlds},} in Society of Photo-Optical Instrumentation Engineers (SPIE) Conference Series, Vol. 9908, Ground-based and Airborne Instrumentation for Astronomy VI, ed. C.~J. {Evans}, L.~{Simard}, \& H.~{Takami}, 99086T, \dodoi{10.1117/12.2233002}

\bibitem[{F. {Kr{\"o}ll} {et~al.}(2026){Kr{\"o}ll}, {Sowmya}, {Shapiro}, {Collier Cameron}, {Witzke}, {Solanki}, {Ribas}, {Shelyag}, {Kopp}, {N{\`e}mec}, \& {Stucki}}]{Florian26}
{Kr{\"o}ll}, F., {Sowmya}, K., {Shapiro}, A.~I., {et~al.} 2026, \bibinfo{title}{{Decoding the Radial Velocity Signatures of Solar Faculae with 3D Magnetohydrodynamic Simulations},} \apj, 1005, 86, \dodoi{10.3847/1538-4357/ae71d1}

\bibitem[{H.-G. {Ludwig} {et~al.}(2009){Ludwig}, {Caffau}, {Steffen}, {Freytag}, {Bonifacio}, \& {Ku{\v{c}}inskas}}]{Ludwigetal2009}
{Ludwig}, H.-G., {Caffau}, E., {Steffen}, M., {et~al.} 2009, \bibinfo{title}{{The CIFIST 3D model atmosphere grid.},} \memsai, 80, 711, \dodoi{10.48550/arXiv.0908.4496}

\bibitem[{Z. {Magic} {et~al.}(2013){Magic}, {Collet}, {Asplund}, {Trampedach}, {Hayek}, {Chiavassa}, {Stein}, \& {Nordlund}}]{Magic2013}
{Magic}, Z., {Collet}, R., {Asplund}, M., {et~al.} 2013, \bibinfo{title}{{The Stagger-grid: A grid of 3D stellar atmosphere models. I. Methods and general properties},} \aap, 557, A26, \dodoi{10.1051/0004-6361/201321274}

\bibitem[{N. {Meunier} {et~al.}(2010){Meunier}, {Desort}, \& {Lagrange}}]{Meunieretal2010}
{Meunier}, N., {Desort}, M., \& {Lagrange}, A.-M. 2010, \bibinfo{title}{{Using the Sun to estimate Earth-like planets detection capabilities . II. Impact of plages},} \aap, 512, A39, \dodoi{10.1051/0004-6361/200913551}

\bibitem[{F. {Pepe} {et~al.}(2002){Pepe}, {Mayor}, {Galland}, {Naef}, {Queloz}, {Santos}, {Udry}, \& {Burnet}}]{Pepe02}
{Pepe}, F., {Mayor}, M., {Galland}, F., {et~al.} 2002, \bibinfo{title}{{The CORALIE survey for southern extra-solar planets VII. Two short-period Saturnian companions to <ASTROBJ>HD 108147</ASTROBJ> and <ASTROBJ>HD 168746</ASTROBJ>},} \aap, 388, 632, \dodoi{10.1051/0004-6361:20020433}

\bibitem[{F. {Pepe} {et~al.}(2021){Pepe}, {Cristiani}, {Rebolo}, {Santos}, {Dekker}, {Cabral}, {Di Marcantonio}, {Figueira}, {Lo Curto}, {Lovis}, {Mayor}, {M{\'e}gevand}, {Molaro}, {Riva}, {Zapatero Osorio}, {Amate}, {Manescau}, {Pasquini}, {Zerbi}, {Adibekyan}, {Abreu}, {Affolter}, {Alibert}, {Aliverti}, {Allart}, {Allende Prieto}, {{\'A}lvarez}, {Alves}, {Avila}, {Baldini}, {Bandy}, {Barros}, {Benz}, {Bianco}, {Borsa}, {Bourrier}, {Bouchy}, {Broeg}, {Calderone}, {Cirami}, {Coelho}, {Conconi}, {Coretti}, {Cumani}, {Cupani}, {D'Odorico}, {Damasso}, {Deiries}, {Delabre}, {Demangeon}, {Dumusque}, {Ehrenreich}, {Faria}, {Fragoso}, {Genolet}, {Genoni}, {G{\'e}nova Santos}, {Gonz{\'a}lez Hern{\'a}ndez}, {Hughes}, {Iwert}, {Kerber}, {Knudstrup}, {Landoni}, {Lavie}, {Lillo-Box}, {Lizon}, {Maire}, {Martins}, {Mehner}, {Micela}, {Modigliani}, {Monteiro}, {Monteiro}, {Moschetti}, {Murphy}, {Nunes}, {Oggioni}, {Oliveira}, {Oshagh}, {Pall{\'e}}, {Pariani}, {Poretti}, {Rasilla}, {Rebord{\~a}o}, {Redaelli}, {Santana
  Tschudi}, {Santin}, {Santos}, {S{\'e}gransan}, {Schmidt}, {Segovia}, {Sosnowska}, {Sozzetti}, {Sousa}, {Span{\`o}}, {Su{\'a}rez Mascare{\~n}o}, {Tabernero}, {Tenegi}, {Udry}, \& {Zanutta}}]{Pepeetal2021}
{Pepe}, F., {Cristiani}, S., {Rebolo}, R., {et~al.} 2021, \bibinfo{title}{{ESPRESSO at VLT. On-sky performance and first results},} \aap, 645, A96, \dodoi{10.1051/0004-6361/202038306}

\bibitem[{N.~E. {Piskunov} {et~al.}(1995){Piskunov}, {Kupka}, {Ryabchikova}, {Weiss}, \& {Jeffery}}]{VALD}
{Piskunov}, N.~E., {Kupka}, F., {Ryabchikova}, T.~A., {Weiss}, W.~W., \& {Jeffery}, C.~S. 1995, \bibinfo{title}{{VALD: The Vienna Atomic Line Data Base.},} \aaps, 112, 525

\bibitem[{H. {Rauer} {et~al.}(2025){Rauer}, {Aerts}, {Cabrera}, {Deleuil}, {Erikson}, {Gizon}, {Goupil}, {Heras}, {Walloschek}, {Lorenzo-Alvarez}, {Marliani}, {Martin-Garcia}, {Mas-Hesse}, {O'Rourke}, {Osborn}, {Pagano}, {Piotto}, {Pollacco}, {Ragazzoni}, {Ramsay}, {Udry}, {Appourchaux}, {Benz}, {Brandeker}, {G{\"u}del}, {Janot-Pacheco}, {Kabath}, {Kjeldsen}, {Min}, {Santos}, {Smith}, {Suarez}, {Werner}, {Aboudan}, {Abreu}, {Acu{\~n}a}, {Adams}, {Adibekyan}, {Affer}, {Agneray}, {Agnor}, {Aguirre B{\o}rsen-Koch}, {Ahmed}, {Aigrain}, {Al-Bahlawan}, {Alcacera Gil}, {Alei}, {Alencar}, {Alexander}, {Alfonso-Garz{\'o}n}, {Alibert}, {Allende Prieto}, {Almeida}, {Alonso Sobrino}, {Altavilla}, {Althaus}, {Alvarez Trujillo}, {Amarsi}, {Ammler-von Eiff}, {Am{\^o}res}, {Andrade}, {Antoniadis-Karnavas}, {Ant{\'o}nio}, {Aparicio del Moral}, {Appolloni}, {Arena}, {Armstrong}, {Aroca Aliaga}, {Asplund}, {Audenaert}, {Auricchio}, {Avelino}, {Baeke}, {Bailli{\'e}}, {Balado}, {Ballber Balaguer{\'o}}, {Balestra}, {Ball},
  {Ballans}, {Ballot}, {Barban}, {Barbary}, {Barbieri}, {Barcel{\'o} Forteza}, {Barker}, {Barklem}, {Barnes}, {Barrado Navascues}, {Barragan}, {Baruteau}, {Basu}, {Baudin}, {Baumeister}, {Bayliss}, {Bazot}, {Beck}, {Belkacem}, {Bellinger}, {Benatti}, {Benomar}, {B{\'e}rard}, {Bergemann}, {Bergomi}, {Bernardo}, {Biazzo}, {Bignamini}, {Bigot}, {Billot}, {Binet}, {Biondi}, {Biondi}, {Birch}, {Bitsch}, {Bluhm Ceballos}, {B{\'o}di}, {Bogn{\'a}r}, {Boisse}, {Bolmont}, {Bonanno}, {Bonavita}, {Bonfanti}, {Bonfils}, {Bonito}, {Bonomo}, {B{\"o}rner}, {Boro Saikia}, {Borreguero Mart{\'\i}n}, {Borsa}, {Borsato}, {Bossini}, {Bouchy}, {Bou{\'e}}, {Boufleur}, {Boumier}, {Bourrier}, {Bowman}, {Bozzo}, {Bradley}, {Bray}, {Bressan}, {Breton}, {Brienza}, {Brito}, {Brogi}, {Brown}, {Brown}, {Brun}, {Bruno}, {Bruns}, {Buchhave}, {Bugnet}, {Buldgen}, {Burgess}, {Busatta}, {Busso}, {Buzasi}, {Caballero}, {Cabral}, {Cabrero Gomez}, {Calderone}, {Cameron}, {Cameron}, {Campante}, {Campos Gestal}, {Canto Martins}, {Cara}, {Carone},
  {Carrasco}, {Casagrande}, {Casewell}, {Cassisi}, {Castellani}, {Castro}, {Catala}, {Catal{\'a}n Fern{\'a}ndez}, {Catelan}, {Cegla}, {Cerruti}, {Cessa}, {Chadid}, {Chaplin}, {Charpinet}, {Chiappini}, {Chiarucci}, {Chiavassa}, {Chinellato}, {Chirulli}, {Christensen-Dalsgaard}, {Church}, {Claret}, {Clarke}, {Claudi}, {Clermont}, {Coelho}, {Coelho}, {Cogato}, {Colom{\'e}}, {Condamin}, {Conde Garc{\'\i}a}, \& {Conseil}}]{Rauer25}
{Rauer}, H., {Aerts}, C., {Cabrera}, J., {et~al.} 2025, \bibinfo{title}{{The PLATO mission},} Experimental Astronomy, 59, 26, \dodoi{10.1007/s10686-025-09985-9}

\bibitem[{M. {Rempel}(2014){Rempel}}]{Rempel14}
{Rempel}, M. 2014, \bibinfo{title}{{Numerical Simulations of Quiet Sun Magnetism: On the Contribution from a Small-scale Dynamo},} \apj, 789, 132, \dodoi{10.1088/0004-637X/789/2/132}

\bibitem[{T. {Ryabchikova} {et~al.}(2015){Ryabchikova}, {Piskunov}, {Kurucz}, {Stempels}, {Heiter}, {Pakhomov}, \& {Barklem}}]{Ryabchikova2015}
{Ryabchikova}, T., {Piskunov}, N., {Kurucz}, R.~L., {et~al.} 2015, \bibinfo{title}{{A major upgrade of the VALD database},} \physscr, 90, 054005, \dodoi{10.1088/0031-8949/90/5/054005}

\bibitem[{C. {Schwab} {et~al.}(2016){Schwab}, {Rakich}, {Gong}, {Mahadevan}, {Halverson}, {Roy}, {Terrien}, {Robertson}, {Hearty}, {Levi}, {Monson}, {Wright}, {McElwain}, {Bender}, {Blake}, {St{\"u}rmer}, {Gurevich}, {Chakraborty}, \& {Ramsey}}]{Schwabetal2016}
{Schwab}, C., {Rakich}, A., {Gong}, Q., {et~al.} 2016, \bibinfo{title}{{Design of NEID, an extreme precision Doppler spectrograph for WIYN},} in Society of Photo-Optical Instrumentation Engineers (SPIE) Conference Series, Vol. 9908, Ground-based and Airborne Instrumentation for Astronomy VI, ed. C.~J. {Evans}, L.~{Simard}, \& H.~{Takami}, 99087H, \dodoi{10.1117/12.2234411}

\bibitem[{N. {Shchukina} \& J. {Trujillo Bueno}(2001){Shchukina} \& {Trujillo Bueno}}]{Shchukina01}
{Shchukina}, N., \& {Trujillo Bueno}, J. 2001, \bibinfo{title}{{The Iron Line Formation Problem in Three-dimensional Hydrodynamic Models of Solar-like Photospheres},} \apj, 550, 970, \dodoi{10.1086/319789}

\bibitem[{K. {Sowmya} {et~al.}(2026){Sowmya}, {Shapiro}, {Vasilyev}, {Witzke}, {Collier Cameron}, \& {Solanki}}]{Sowmya25}
{Sowmya}, K., {Shapiro}, A.~I., {Vasilyev}, V., {et~al.} 2026, \bibinfo{title}{{Sensitivity of Spectral Lines to Granulation: The Sun},} \apj, 1003, 28, \dodoi{10.3847/1538-4357/ae6102}

\bibitem[{S. van~der Walt {et~al.}(2011)van~der Walt, Colbert, \& Varoquaux}]{numpy}
van~der Walt, S., Colbert, S.~C., \& Varoquaux, G. 2011, \bibinfo{title}{The NumPy Array: A Structure for Efficient Numerical Computation,} Computing in Science Engineering, 13, 22, \dodoi{10.1109/MCSE.2011.37}

\bibitem[{V. {Vasilyev} {et~al.}(2026){Vasilyev}, {Sowmya}, {Shapiro}, {Kostogryz}, {Vukadinovic}, {Witzke}, {Bhatia}, {Cameron}, {Gizon}, \& {Solanki}}]{Valera26}
{Vasilyev}, V., {Sowmya}, K., {Shapiro}, A.~I., {et~al.} 2026, \bibinfo{title}{{Sensitivity of Spectral Lines to Granulation: From the Sun to K-type Stars},} \apj, 1004, 206, \dodoi{10.3847/1538-4357/ae6e33}

\bibitem[{P. Virtanen {et~al.}(2020)Virtanen, Gommers, Oliphant, Haberland, Reddy, Cournapeau, Burovski, Peterson, Weckesser, Bright, {van der Walt}, Brett, Wilson, Millman, Mayorov, Nelson, Jones, Kern, Larson, Carey, Polat, Feng, Moore, {VanderPlas}, Laxalde, Perktold, Cimrman, Henriksen, Quintero, Harris, Archibald, Ribeiro, Pedregosa, {van Mulbregt}, \& {SciPy 1.0 Contributors}}]{scipy}
Virtanen, P., Gommers, R., Oliphant, T.~E., {et~al.} 2020, \bibinfo{title}{{{SciPy} 1.0: Fundamental Algorithms for Scientific Computing in Python},} Nature Methods, 17, 261, \dodoi{10.1038/s41592-019-0686-2}

\bibitem[{A. {V{\"o}gler} {et~al.}(2004){V{\"o}gler}, {Bruls}, \& {Sch{\"u}ssler}}]{Voegleretal2004}
{V{\"o}gler}, A., {Bruls}, J.~H.~M.~J., \& {Sch{\"u}ssler}, M. 2004, \bibinfo{title}{{Approximations for non-grey radiative transfer in numerical simulations of the solar photosphere},} \aap, 421, 741, \dodoi{10.1051/0004-6361:20047043}

\bibitem[{A. {V{\"o}gler} {et~al.}(2005){V{\"o}gler}, {Shelyag}, {Sch{\"u}ssler}, {Cattaneo}, {Emonet}, \& {Linde}}]{Voegleretal2005}
{V{\"o}gler}, A., {Shelyag}, S., {Sch{\"u}ssler}, M., {et~al.} 2005, \bibinfo{title}{{Simulations of magneto-convection in the solar photosphere. Equations, methods, and results of the MURaM code},} \aap, 429, 335, \dodoi{10.1051/0004-6361:20041507}

\bibitem[{V. {Witzke} {et~al.}(2023){Witzke}, {Duehnen}, {Shapiro}, {Przybylski}, {Bhatia}, {Cameron}, \& {Solanki}}]{Witzke23}
{Witzke}, V., {Duehnen}, H.~B., {Shapiro}, A.~I., {et~al.} 2023, \bibinfo{title}{{Small-scale dynamo in cool stars. II. The effect of metallicity},} \aap, 669, A157, \dodoi{10.1051/0004-6361/202244771}

\bibitem[{V. {Witzke} {et~al.}(2021){Witzke}, {Shapiro}, {Cernetic}, {Tagirov}, {Kostogryz}, {Anusha}, {Unruh}, {Solanki}, \& {Kurucz}}]{Witzke21}
{Witzke}, V., {Shapiro}, A.~I., {Cernetic}, M., {et~al.} 2021, \bibinfo{title}{{MPS-ATLAS: A fast all-in-one code for synthesising stellar spectra},} \aap, 653, A65, \dodoi{10.1051/0004-6361/202140275}

\bibitem[{V. {Witzke} {et~al.}(2024){Witzke}, {Shapiro}, {Kostogryz}, {Mauviard}, {Bhatia}, {Cameron}, {Gizon}, {Przybylski}, {Solanki}, {Unruh}, \& {Yue}}]{Witzke24}
{Witzke}, V., {Shapiro}, A.~I., {Kostogryz}, N.~M., {et~al.} 2024, \bibinfo{title}{{Testing MURaM and MPS-ATLAS against the quiet solar spectrum},} \aap, 681, A81, \dodoi{10.1051/0004-6361/202346099}

\end{thebibliography}
\bibliographystyle{aasjournalv7}

\appendix

\section{Atmospheric stratification comparison}
\label{sec:app_atmos}

\begin{figure}[h!]
    \centering
    \includegraphics[width=\linewidth]{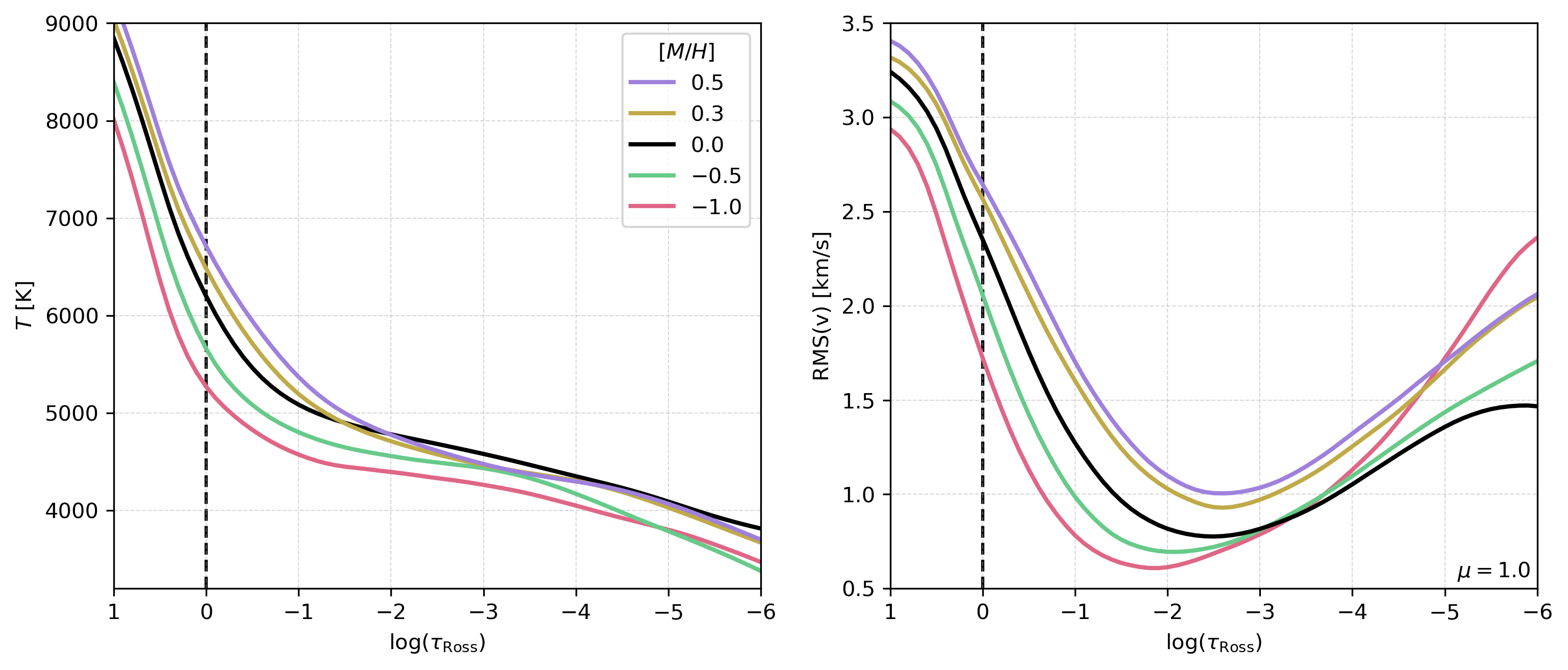}
    \caption{Stratification of the mean temperature (left panel) and vertical velocity RMS (right panel) for considered metallicity cases at the disc center ($\mu=1$). The color coding is the same as in Figure\,\ref{fig:proxy_comparison}. The vertical dashed line indicates the stellar surface ($\log(\tau_\mathrm{Ross})=0$).}
    \label{fig:avgT_RMSvz}
\end{figure}

Figure\,\ref{fig:avgT_RMSvz} shows the mean temperature and the RMS of the vertical velocity in a single snapshot for each considered metallicity. The metal-rich models show increased temperature and larger velocity variation, whereas metal-poor models have lower temperature and smaller velocity variation compared to the solar $M/H=0.0$. The largest deviation in this ordering with metallicity is observed for the velocity variation for $M/H=-1.0$ metallicity in atmospheric layers above $\log(\tau_\mathrm{Ross})=-2$.

\end{document}